# Magnetic-field tuned anisotropic quantum phase transition in the distorted kagome antiferromagnet Nd$_3$BWO$_9$

Fangyuan Song[1*], Han Ge[2*], Andi Liu[1], Yuqi Qin[1], Yuyan Han[3], Langsheng Ling[3], Songliu Yuan[1], Zhongwen Ouyang[1], Jieming Sheng[2,4+], Liusuo Wu[2,5‡], Zhaoming Tian[1,6#]

[1] School of Physics and Wuhan National High Magnetic Field Center, Huazhong University of Science and Technology, Wuhan, 430074, China

[2] Department of Physics, Southern University Science and Technology, Shenzhen, 518055, China

[3] Anhui Province Key Laboratory of Condensed Matter Physics at Extreme Conditions, High Magnetic Field Laboratory, Chinese Academy of Sciences, Hefei, 230031, China

[4] Spallation Neutron Source Science Center, Dongguan, 523803, China

[5] Shenzhen Key Laboratory of Advanced Quantum Functional Materials and Devices, Southern University of Science and Technology, Shenzhen, 518055, China

[6] Shenzhen Huazhong University of Science and Technology research institute, Shenzhen, 518057, China

**ABSTRACT:**

Rare-earth (RE) kagome-lattice magnets offer an excellent platform to discover the novel magnetic phase as well as quantum phase transition tuned by non-thermal control parameters, while the experimental realizations remain largely unexplored. Here, we report the discovery of magnetic-field ($B$)-induced anisotropic quantum phase transition in a distorted kagome antiferromagnet Nd$_3$BWO$_9$ with $T_N$~0.32 K. The isothermal magnetizations at 0.05 K exhibit the spin-flop like metamagnetic crossover behaviors with different fractional magnetization anomalies for $B$ perpendicular ($B$ // $c$-axis) and parallel ($B$ // $a^*$-axis) to the kagome plane, respectively. In combination with the thermodynamic measurements, the field-temperature ($B$-$T$) phase diagrams for both field directions are constructed and that reveal the existence of several field-induced magnetic states. Along the $c$-axis, a proximate quantum bicritical point is observed near the metamagnetic crossover, which separates the low-field antiferromagnetic (AFM) phase and the intermediate AFM phase. While, for $B$ // $a^*$, another intermediate magnetic phase (IAFM2) appears between the low-field AFM phase and intermediate AFM (IAFM1) phase, giving rise to a tetracritical point. These results support the anisotropic field-induced metamagnetic quantum criticalities in Nd$_3$BWO$_9$, making it as a rare kagome antiferromagnet to investigate the quantum multi-criticality driven by spin frustration.

---

[*] These authors contributed equally to this work.
[+] Corresponding author: shengjm@sustech.edu.cn
[‡] Corresponding author: wuls@sustech.edu.cn
[#] Corresponding author: tianzhaoming@hust.edu.cn

## ◼ INTRODUCTION

Rare-earth (RE)-based geometrically frustrated magnets (GFMs) have attracted great attention in searching for the exotic magnetic phases of matter and fractionalized spin excitations. Compared to the relatively well-studied 3$d$ transition-metal systems, the materials containing 4$f$ RE ions have strong spin-orbital coupling (SOC), large single-ion anisotropy, dipolar exchange interaction and rich diversity of spin-types. These complex interactions and inherent lattice frustration render RE-based GFMs as candidates to discover novel quantum states beyond the 3$d$-based ones including Kitaev quantum spin-liquid (QSL) [1-3], quantum spin ice [4-6] and quantum dipolar phase [7]. Moreover, the relatively smaller exchange interactions between localized 4$f$ moments allow an easy tunability of spin states through the disorder-free control parameters accessible in the laboratory such as external field or pressure.

As one type of spin-frustrated systems, kagome-lattice magnets with different anisotropy between the limits of isotropic Heisenberg type and Ising-like model, have attracted a lot of theoretical and experimental investigations on exploring the exotic spin states [6,8-10]. Compared to the extensive experimental studies on transition-metal kagome materials, RE-based kagome magnets with considerably weaker exchange interactions have been less investigated. So far, the studied materials can be divided into two categories: one is the intermetallic compound containing conductive electrons, the other is the insulating oxide with magnetism fully determined by the superexchange interactions of 4$f$ local moments. For the former, typical examples include REAgGe [11,12], RE$_3$Ru$_4$Al$_{12}$ [13,14] and REPdAl [15,16], owing to the hybridization between conductive electrons and 4$f$ local moments, unconventional magnetic ground state can be generated by adjusting the Ruderman-Kittel-Kasuya-Yosida (RKKY) interaction as well as the Kondo effect between local moments [17]. For the latter, the available experimental studies are only limited to a few systems including the two-dimensional (2D) kagome RE$_3$Sb$_3$M$_2$O$_{14}$ (M = Zn, Mg) [6,18-21] and RE$_3$Ga$_5$SiO$_{14}$ (RE = Pr, Nd) [22-25] families, and their magnetic ground states remain to be fully understood. For example, in distorted kagome Nd$_3$Ga$_5$SiO$_{14}$, the magnetic ground state detected by different experimental techniques seems to be confusing, where muon spin relaxation ($\mu$SR) and nuclear magnetic resonance (NMR) results indicate a characteristic of spin liquid with spin fluctuating down to 60 mK [23] while the measurements of magneto-optical spectroscopy and neutron scattering support a single ion behavior instead of spin liquid state down to 100 mK [26,27]. For RE$_3$Sb$_3$M$_2$O$_{14}$ compounds, the unavoidable antisite disorder between magnetic RE$^{3+}$ and nonmagnetic M$^{2+}$ ions complicates the clarification on its intrinsic magnetic ground state [28,29], and the lacking of single crystals so far hinders further detailed investigation on their exotic magnetic behaviors especially the ones related to crystalline directions.

Despite the exploration of novel magnetic ground state, magnetic-field ($B$) induced quantum phase transition (QPT) between two different ground states as a consequence of quantum correlation also attracts strong interest [30], where quantum spin fluctuations can stabilize an exotic magnetic phase or a particular spin configuration exhibiting fractional magnetization in a certain field regime. In this direction, most studies on RE-based kagome materials have carried out the heavy fermion systems containing $Ce^{3+}$ ($4f^1$) or $Yb^{3+}$($4f^{13}$) ions [17, 31-36]. In these systems, both RKKY interaction and Kondo coupling can play important roles on the QPT, and consequently, they enrich the magnetic phase diagrams and in certain conditions lead to the appearance of unconventional quantum criticalities, such as the quadrupolar order and non-Fermi liquid near a quantum critical point, etc. [33-36]. On the other side, the competing multiple interactions complicate the underlying mechanism as an obstacle for its clarification of QPT behaviors. From this viewpoint, the insulating counterparts are more attractive to unveil the QPT dictated by the kagome-lattice topology of $RE^{3+}$ moments. Conversely, their physical clarification will help to understand the QPT of kagome-lattice heavy fermion systems in case of the reintroduced conductive electrons, but the experimental work is rare due to the lack of an appropriate model system.

$Nd_3BWO_9$ belongs to one series of RE-based kagome compounds $RE_3BWO_9$ (RE=Pr,Nd, Gd-Ho) crystallized in a hexagonal structure with space group $P6_3$ [37,38], where magnetic $Nd^{3+}$ ions occupy the distorted kagome lattices in the *ab*-plane stacking in the "ABAB"-type fashion along *c*-axis [see Figs.1(a,b)]. Thus, it provides a candidate to study the QPT phenomena in RE-based kagome insulator. In this work, through the low temperature magnetization, specific heat and magnetocaloric effect (MCE) measurements down to 0.05 K, we experimentally uncover the field-induced anisotropic QPTs in $Nd_3BWO_9$ antiferromagnet. Below $T_N$~0.32 K, the field-induced metamagnetic (MM) crossover behaviors are identified with different fractional magnetizations for $B$ applied perpendicular ($B // c$) and parallel ($B // a^*$) to the kagome plane, respectively. Moreover, the constructed $B$-$T$ phase diagrams indicate the anisotropic multiple critical behaviors, where the metamagnetic bicritical point ($B // c$) and tetracritical point ($B // a^*$) are detected in the phase diagrams.

## II. EXPERIMENTAL DETAILS

Single crystals of $Nd_3BWO_9$ were grown by a high-temperature flux method similar to the previous report [39]. The as-grown $Nd_3BWO_9$ single crystal is shown in Fig. 1(d). Also, the $La_3BWO_9$ samples were synthesized using as the nonmagnetic reference of specific heat. The crystal structure was characterized by x-ray diffraction (XRD) diffraction using a Bruker D8 x-ray machine with Cu $K\alpha$ radiation ($\lambda$ = 1.5418 Å) using the powders from the crashed crystals. Also, the crystal structure was determined by single-crystal X-ray diffraction (SXRD) with the diffraction data collected by a Kappa Apex2 CCD diffractometer (Bruker) using graphite-monochromated Mo $K_\alpha$ radiation ($\lambda$ = 0.71073 Å). The obtained crystal data from the structure refinements are provided in Table S1 and S2 in supplemental materials [40].

Temperature-dependent *dc* susceptibility $\chi$(T) and isothermal magnetization $M(B)$ measurements were performed on the Quantum Design superconducting quantum interference device (SQUID) magnetometer and commercial Physical Property Measurement System (PPMS) for temperature range of 1.8-300 K. For the lower temperatures (0.05≤$T$≤4 K), magnetic measurements were carried out using a Hall sensor magnetometer integrated with a dilution insert of the PPMS. The specific heat and magnetocaloric effect (MCE) were measured down to 0.06 K in PPMS equipped with a dilution refrigeration system using the heat capacity option. Heat-capacity measurements were carried out under constant field by the thermal relaxation method, and MCE were measured under quasi-adiabatic conditions using a constant field sweeping rate of 20 Oe/s between 0 to 3 T. The thermometer resistance of the heat capacity thermometer was measured and converted to temperatures, and then the MCE values can be derived.

## III. RESULTS AND DISCUSSIONS

### A. Low temperature magnetic ground state

The refined crystal structure of $Nd_3BWO_9$ fits the hexagonal $P6_3$ space group with lattice parameters $a=b$=8.695(2) Å and $c$=5.485(1) Å, in consistent with the previous results [37]. The detailed information on atomic coordinates and bond distances from the single crystal refinements are listed in Table S1 and S2. In the unit cell, magnetic $Nd^{3+}$ ions are located on the distorted dodecahedron site surrounded by 8 oxygens with completely different Nd-O bond distances [see Fig. 1(c)]. Through the corner- or edge-sharing interconnections with $WO_6$ octahedra and planar $BO_3$ groups, $NdO_8$ polyhedra are linked to form a distorted "breathing" kagome network within the *ab*-plane. The side lengths of the $Nd^{3+}$ equilateral triangles are $r_1^{intra}$=4.248 Å and $r_2^{intra}$=4.929 Å, which are slightly larger than the interlayer's separation $r^{inter}$=3.950 Å. Using the side length of $Nd^{3+}$ equilateral triangles, the distortion value of breathing kagome lattice is evaluated to $\frac{r_2^{intra}-r_1^{intra}}{r_2^{intra}+r_1^{intra}} \times 100\% \sim 7.4\%$ and that is larger than the value ~2.6% in distorted kagome $Nd_3Ga_5SiO_{14}$ [22]. From the structure viewpoint, another important feature is the large difference of ionic radii and distinct local oxygen coordination environments between magnetic $Nd^{3+}$ and nonmagnetic $B^{3+}/W^{6+}$ cations, which can minimize the antisite disorder. Compared to the kagome $RE_3Sb_3M_2O_{14}$ (M=Zn,Mg) systems, the antisite disorder between $RE^{3+}$ and $M^{2+}$ ions impede the clarification of intrinsic magnetic ground state [28,29], here the weak disorder effect lets $Nd_3BWO_9$ as clean system to study the intrinsic kagome magnetism.

The isothermal magnetization $M(B)$ measurements at 2 K on $Nd_3BWO_9$ single crystals have been carried out for $B$ along the $c$, $a^*$ and $b^*$ axes, the definition of axis direction is schematically shown in the inset of Fig. 1(e). The $M(B)$ results indicate a magnetic anisotropy, the $c$-axis magnetization has the largest saturated magnetization $M_s \sim 1.71 \mu_B$/Nd. As $B$ rotates within the $ab$-plane [see Fig. 1(f)], angle-dependent magnetizations show a six-fold symmetry, accordingly, $a^*$ and $b^*$ axis is along the direction exhibiting the maximum and minimum moment values, in this definition, $a^*$ axis tilts from the crystallographic $b$ axis by a canting angle $\sim 7°$ within the $ab$-plane. Fig. 1(g) shows the inverse magnetic susceptibilities $\chi^{-1}(T)$ along $c$ and $a^*$ axes, the high temperature ($T > 150$ K) Curie-Weiss (CW) fits give the CW temperature $\theta_{cw}$ = -39.7 K ($B // c$) and $\theta_{cw}$= -38.2 K ($B // a^*$) and effective moment $\mu_{eff}$ = 3.59 $\mu_B$ ($B // c$) and $\mu_{eff}$= 3.68 $\mu_B$ ($B // a^*$), respectively. The obtained $\mu_{eff}$ are close to the free-ion value of 3.62$\mu_B$ for $Nd^{3+}$ ($4f^3$, $J$=9/2) ion, but large value of $|\theta_{cw}|$ doesn't indicate strong antiferromagnetic (AFM) interactions between $Nd^{3+}$ moments due to the thermal population of excited crystal electric field (CEF) levels [37,41]. Thereby, the CW fitting is also performed at low temperatures $T<$ 10 K [see the inset of Fig. 1(g)], which gives effective moments $\mu_{eff}$ = 2.92 $\mu_B$ ($B // c$) and $\mu_{eff}$= 2.85 $\mu_B$ ($B // a^*$). The low temperature fitted $\theta_{cw}$ = -0.97 K ($B // c$) and $\theta_{cw}$= -1.31 K ($B // a^*$) reveal the dominant AFM interactions between $Nd^{3+}$ moments on the order of a kelvin.

To determine the magnetic ground state, zero-field specific heat $C_p(T)$ of $Nd_3BWO_9$ with temperature down to 0.1 K is shown in Fig. 2(a). The typical feature is the broad maximum centered at $T_s \sim 0.85$ K and λ-like peak at $T_N \sim 0.32$ K, the latter signals the onset of AFM ordered state. Below 0.15 K, the upturn of $C_p(T)$ comes from the contribution of hyperfine field enhanced by nuclear Schottky anomaly ($C_{Nuc}$), the grey dashed line in Fig. 2(a) shows the estimation by $C_{Nuc} \propto 1/T^2$ in a first approach. After further subtraction of the lattice contribution ($C_{Latt}$) using the nonmagnetic analogue $La_3BWO_9$ as background, the magnetic specific heat $C_M(T)$ of $Nd_3BWO_9$ is obtained as plotted in Fig. 2(b). Besides the two peaks at $T_N$ and $T_{sr}$, another high-$T$ broad hump maximized at ~72 K is visible, which is in associated with the Schottky anomaly with CEF energy splitting of the $J$=9/2 multiplet of $Nd^{3+}$ ions. From the fit to $C_M(T)$ data by the two-level model, the first CEF excited level is estimated to be ~13 meV, this large energy gap ensures the well separated Kramers doublet ground state from the 1st excited state. Thus, $Nd_3BWO_9$ can be well considered as an effective $S_{eff}$=1/2 moment system in the low temperature and low field regimes.

The magnetic entropy $S_M(T)$ evaluated by integrating $C_{mag}/T$ over temperature is shown in Fig. 2(b). As increased $T$, $S_M(T)$ reaches a saturation value $\sim R\ln2$ at 10 K as expected for the doublet ground state. While, the release of magnetic entropy at $T_N$ only reaches $\sim 0.16 R\ln2$, this implies the strong spin fluctuation in temperature range of $T_N<T<10$ K and which prevents a total ordering of $Nd^{3+}$ moments, then the broad peak at $T_{sr}$ is ascribed to the short-range spin correlation. Under applying magnetic field ($B$ //$c$), both $T_N$ and $T_{sr}$ show the nonmonotonic temperature dependencies as guided by the dashed lines in Fig. 2(c). As increased $B \geq 1$ T, $T_N$ is suppressed to below 0.1 K and the system enters the fully polarized (FP) state.

## B. Field-induced metamagnetism and fractional magnetizations

Figures. 3(a)-3(f) show the isothermal $M(B)$ and its field derivative d$M$/d$B$ curves at low temperatures of 0.05 K $\leq T \leq$ 2 K for $B$ along the three typical axes. At $T = 50$ mK, all $M(B)$ curves reach the saturations at small field $B_s<$ 1.2 T with magnetic moments $M_{S,c}=1.77\mu_B$/Nd, $M_{S,a^*}=1.50\mu_B$/Nd and $M_{S,b^*}=1.32\mu_B$/Nd for $B$ along $c$, $a^*$ and $b^*$, respectively. Comparing the $M(B)$ and d$M$/d$B$ curves, we can identify an obvious magnetic anisotropy. For $B$ // $c$, the $M(B)$ curves exhibit the spin-flop-like metamagnetic (MM) transition with an inflection point between the critical fields $B_{c1}\sim$0.54 T and $B_{c2}\sim$0.65 T, where $B_{c1}$ and $B_{c2}$ are determined by the peak of the d$M$/d$B$ curves [see Fig. 3(b)]. Following this metamagnetic transition, magnetization shows a weak step at 1/3 $M_{S,c}$, then it increases steeply and approaches its saturation at $B > B_s$. By contrast, for $B$ // $a^*$, two successive metamagnetic transitions are detected with critical fields $B_{a^*1}\sim$0.49T and $B_{a^*2}\sim$0.63 T, where the $M(B)$ curves show two fractional magnetization anomalies at $\sim$1/6 $M_{s,a^*}$ and $\sim$1/4$M_{s,a^*}$ [see Fig. 3(c) and 3(d)]. Also, a similar 1/4 magnetization anomaly is observed along the $b^*$ axis. These results demonstrate the anisotropic metamagnetism evidenced by the 1/3- and 1/4- fractional magnetizations for $B$ perpendicular and parallel to the kagome plane, respectively. Here, it is noteworthy that the in-plane 1/4$M_s$ in $Nd_3BWO_9$ is distinct from the 1/3$M_s$ in uniform kagome AFMs [42-44]. For the emergence of fractional magnetization under magnetic field in kagome lattice magnets, recent theoretical studies have revealed the important roles of lattice distortion on the diversity of fractional magnetizations where different spin configurations can be stabilized by tunable spin-lattice coupling and spatial exchange anisotropy [45,46]. Herein, the in-plane 1/4 fractional magnetization can be ascribed to the anisotropic planar magnetic interactions inherent into the distorted breathing kagome lattice of $Nd^{3+}$ moments, leading to a magnetic structure with inter-spin angle deviating from 120° under magnetic field. This is in accordance with the recent theoretical analysis on the RE-kagome spin system [46], where the 1/6$M_S$ and 1/4$M_S$ can be induced by the lattice deformation under in-plane field.

As increased temperature, the d$M$/d$B$ peaks near the metamagnetic crossover become weak and then disappear at $T > T_N$, while the anomalies at ~$B_s$ survive up to $T > 1$ K. To get more information on the low-$T$ magnetic behavior, magnetic susceptibilities $\chi(T)$ in temperature range of 0.04 - 3 K are presented for $B$ along $c$ and $a^*$ axis in Figs. 4(a,b). The $\chi(T)$ taken at $B$ = 0.2 T exhibits a rounded maximum at $T_{sr}$~1.2 K, signifying the short-range spin correlation usually observed in the low-dimensional spin frustrated systems [3, 47]. Additionally, $T_N$ can be determined by the maxima in the $\chi(T)$ curves at intermediate fields (0.4 T< $B$ < 0.8 T) denoted by arrows in Figs. 4(a,b). As increased $B$, it moves to low temperature and becomes undistinguishable for $B > B_s$ when it enters into the FP magnetic state.

### C. Field-temperature magnetic phase diagrams

To establish the phase boundaries of $B$-induced phase transitions in $Nd_3BWO_9$, the isothermal field scans on specific heat $C(B)$ were measured at different temperatures for $B$ along $c$ and $a^*$ axis, the representative $C(B)/T$ data are plotted in Figs. 5(a)-5(d). Along both directions, two anomalous peaks can be identified near the critical fields as field across the metamagnetic transition in the $C(B)/T$ curves at temperatures 0.1 K <$T$<$T_N$, which are separated from the low-$T$ single peak at $T$ < 0.1 K ($B // c$) and $T$ = 0.12 K ($B // a^*$) [see Figs. 5(a) and 5(c)]. Upon increasing $T$, these two critical fields shift in opposite directions, then the "double hump"-shaped anomaly flattens into a "wing"-shape when $T$ is close to $T_N$, as exemplarily shown for the data of $T$ = 0.3 K [see Figs. 5(b) and 5(d)]. At high field sides, another anomaly located at ~$B_s$ is visible for $T \leq 0.2$ K, and it splits into two broad peaks whose magnitudes gradually decrease with increased $T$. Moreover, these two peaks are extended well above $T_N$ characterizing a field-induced magnetic crossover behavior, the peak positions ($B_{s1,c}$,$B_{s2,c}$ for $B// c$ and $B_{s1,a^*}$,$B_{s2,a^*}$ for $B//$ a*) are used to determine the phase boundaries of magnetic crossover.

By combining all the available susceptibility, magnetization and specific heat data, we construct the $B$-$T$ phase diagrams based on the critical fields and ordering temperatures. The contour plots using the $C_M(B)$ data for $B // c$ and $B //a^*$ are shown in Figs. 6(a,b), respectively. For $B // c$, two critical points can be identified in the phase diagram: one is the bicritical point (BCP) as observed in anisotropic AFMs with metamagnetic transition [48,49], at this point ($B_{BCP}$ ~0.61±0.02 T,$T_{BCP}$ ~ 0.03±0.02 K) two critical lines labelled by $T_{N,c1}(B)$ and $T_{N,c2}(B)$ converge; another is a tricritical point (TCP) at $B_{s,c}$~1.03±0.02 T and $T_{TCP}\rightarrow$ 0.18±0.02 K, where the PM and intermediate AFM (IAFM) phase and FP phase meet together. A careful inspection on the location of the BCP reveals that $T_{BCP}$ is quite close to zero Kelvin, and this is supported by the low-$T$ $M(B)$ curves since the BCP should reside at field between $B_{c1}$~0.54 T and $B_{c2}$~0.65 T and temperature below the base temperature of 50 mK (far lower than $T_N$) [ see Fig. 3(b) ]. Thus, it realizes a proximate quantum bicritical point (QBCP) [35,50], around which the observed dip behavior of $T(B)$ curve [see Fig. 7(a)] supports the presence of large quantum fluctuation as discussed later. In this case, the spin-flop line below the BCP shrinks to a single point at $T = 0$ K. On the high-field side, below the TCP, a nearly vertical critical line separating the IAFM and FP phases is observed. At elevated temperatures ($T > T_{TCP}$), it splits into two critical lines forming a $V$-shaped critical regime [ see Fig. 6(a) ]. Additionally, we can find a dome-shaped phase region mapped by $B_{c2}$ and $B_{s,c}$, inside which the 1/3 $M_{s,c}$ plateau emerges. In all, the above $c$-axis phase diagram of $Nd_3BWO_9$ highlights the realization of the quantum bicriticality previously reported in the distorted kagome metals YbAgGe [34] and CeRhSn [35] but not yet in insulating counterparts. In the above kagome metallic systems, the occurrence of QBCP is proposed to be driven by two controlling parameters with dedicate strengths [34,35,51]: one is the ratio of Kondo coupling to RKKY interaction and another is quantum fluctuation by spin frustration, the tunability on both parameters can lead to the appearance of QBCP. In $Nd_3BWO_9$, due to the absence of conductive electrons, the observed quantum bicriticality should be related to the large quantum fluctuation enhanced by magnetic frustration, its existence is indicated by the low temperature $C_M(T)$ and $S_M(T)$ results with only ~16% of magnetic entropy released below $T_N$.

The phase diagrams of $B // a^*$ are shown in Fig. 6(b). Despite the paramagnetic (PM) regime, four magnetic ordered phases can be identified: the low-field AFM ($L$-AFM) phase below $B_{a^*1}$~0.49T, the 1st intermediate magnetic (IAFM1) phase in a narrow field range between $B_{a^*1}$~0.49 T and $B_{a^*2}$~0.68 T, the 2nd intermediate magnetic (IAFM2) phase ended at $B_{s,a^*}$ and high-field FP phase at $B > B_{s,a^*}$. Between the IAFM2 and FP phases, a QCP is identified at $B_{s,a^*}$~1.16±0.03 T and $T_{CP} \to 0$ K, above that there exists a phase region between the phase boundaries of magnetic crossover as denoted by the dashed lines. Compared to $B // c$, the most distinct difference happens near the metamagnetic crossover, the BCP turns out to be a tetracritical point (TP) at which four phase critical lines meet together [52,53]. The TP is located at $B_{TP}$~0.51±0.02T and $T_{TP}$~0.13 ±0.02 K, above that two order-disorder transition lines [$T_{N,a^*1}(B)$ and $T_{N,a^*2}(B)$] separate the PM phase between the $L$-AFM and IAFM2 phases. Below the TP, another two critical lines separate the IAFM1 phase from the $L$-AFM and IAFM2 phases, this is different from the occurrence of spin-flop transition line in the bicritical phase diagram [48,54]. The formation of novel IAFM1 phase is in accordance with the $a^*$-axis $M(B)$ curves at 50 mK, which display two-stage fractional magnetization steps in contrast to the single $1/3M_S$ anomaly for $B // c$ [further comparison is shown in Fig. S2]. Correspondingly, the $1/6M_{S,a^*}$ and $1/4M_{S,a^*}$ anomalies occur inside the IAFM1 and IAFM2 phases in the phase diagrams, respectively.

To characterize the metamagnetic critical behaviors, the phase boundaries are analysed by a power law with the form $T_N(B) \propto |B-B_c|^\phi$, where $B_c$ is the critical field and $\phi$ is the critical exponent. Along the c-axis, by fixing $B_c = B_{BCP} = 0.61(1)$ T and $T_c = 0$ K, the fits to $T_{N,c1}(B)$ and $T_{N,c2}(B)$ yield two critical exponents $\phi_{c1} = 0.52(2)$ and $\phi_{c2} = 0.61(2)$, signifying a change of spin symmetry or dimensionality across the metamagnetic BCP [53]. Also, the fits to $T_{N,c1}(B)$ and $T_{N,c2}(B)$ with the constraint of $\phi_{c1} = \phi_{c2} = \phi_c$ is possible which gives $\phi_c = 0.55(2)$. For $B // a^*$, the corresponding estimation with $B_c = B_{TP} = 0.61(1)$ T gives the critical exponent $\phi_{a^*} = 0.63(2)$ [see the dashed lines in Fig. 6(b)], and this exponent value is coordinated by the fitted value $\phi_{a^*,s} = 0.65(2)$ near the high-field QCP using $B_c = B_{s,a^*} \sim 1.16$ and $T_{CP} = 0$ K. Along these two directions, different critical exponents suggest the anisotropic scaling behaviors, the exponent $\phi_c = 0.55(2)$ for $B // c$ is better agreement with the 3-dimensional (3D) Ising anisotropy with $\phi = 1/2$ [55], while the $\phi_{a^*} = 0.63(2)$ along the $a^*$-axis coincides with the expected value $\phi = 2/3$ of the 3D XY universality as reported in the spin-dimerized $BaCuSi_2O_6$ [56] and spin-frustrated $Cs_2CuCl_4$ AFMs [57].

### D. Thermodynamic evidence of multiple quantum criticality

To get more experimental evidences on the quantum criticality, we carried out the magnetocaloric effect (MCE) measurements under quasi-adiabatic conditions, which can reflect the sample-temperature change induced by the evolution of magnetic entropy under field sweeping. The collected $T(B)$ data measured at different initial temperatures ($T_{ini}$) for $B // c$-axis and $B // a^*$-axes are shown in Figs. 7(a) and 7(c), respectively. Along the c-axis, $T(B)$ curves exhibit dramatic temperature drops as approaching the CPs. For $T_{ini} < T_N$, we can observe a clear temperature dip-like feature near the critical fields $B_{BCP} \sim 0.61$ T and $B_{TCP} \sim 1.03$ T, where the accumulation of magnetic entropy $S_M(B)$ reaches its maximum. Below the TCP, the 1st order-type phase transition for $T_{ini} < \sim 0.15$ K is indicated by an asymmetric peak-like feature of $T(B)$ curves due to the consequence of entropy discontinuity. For $B // a^*$, in temperature regimes 0.12 K $\leq T_{ini} \leq$ 0.5 K, the $T(B)$ curves also exhibit similar temperature drop behaviour close to the TP. As $T_{ini} < 0.1$ K, the temperature dip-like feature is gradually suppressed reflecting the reduced accumulation of $S_M(B)$ at field around $B_{TP}$. At $T_{ini} = 0.06$ K, a pronounced temperature peak appears at ~0.56 T well below the TP, and around which field the $M(B)$ curves show a $1/6 M_{S,a^*}$ anomaly [see Fig. 3(c) and 3(d)].

The magnetic Grüneisen parameter derived from the adiabatic MCE $\Gamma_B(B) = 1/T(dT/dB)$ has been shown to be a suitable probe to identify the QCPs [58]. Figs. 7(b) and 7(d) show the obtained $\Gamma_B(B)$ results for $B // c$ and $B // a^*$, respectively. Along both field directions, as approaching the CPs, the $\Gamma_B(B)$ curves exhibit the singularities with sign change in accordance with the criticality-enhanced MCE [58-60]. To illustrate the multiple quantum criticalities, the *B-T* phase diagrams are mapped out by the contour plots of the $\Gamma_B(B)$ data, where the phase boundaries are determined by the peak positions of d*T*/d*B* curves shown in Figs. 7(b) and 7(d). As seen, the obtained overall phase diagrams and extracted phase boundaries agree well with the ones by the specific heat and magnetization data. In the *c*-axis phase diagram, the position of metamagnetic BCP from the linear extrapolation of $B_{c1}$ and $B_{c2}$ goes to zero temperature in consistent with the realization of QBCP. A peak of $\Gamma_B(B)$ centered at ~1.1 T and ~0.18 K suggests a TCP, and the location of TCP is consistent with the determination by the fields $B_{s1,c}$ and $B_{s2,c}$ of magnetic crossover as shown in Fig. S3. By contrast, for *B* // *a**, the high-field CP resides at zero temperature and around which there exists a quantum critical regime at finite temperatures, the metamagnetic TP appears at finite temperature, below which an enclosed IAFM1 phase region occurs.

## IV. DISCUSSIONS AND CONCLUSIONS

It is interesting to compare the magnetic behaviors of $Nd_3BWO_9$ with ones observed in other RE-based kagome lattice systems. For its magnetic ground state, the coexistence of long-range magnetic order and large spin fluctuation is quite different from the kagome lattice $Nd_3Sb_3Mg_2O_{14}$ with a single magnetic phase transition at $T_N$~0.54 K [19,21] and distorted kagome $Nd_3Ga_5SiO_{14}$ with a spin disorder ground state [23,27]. In terms of field-induced QPT, it is compared with the kagome lattice heavy fermion systems with hexagonal ZrNiAl-type crystal structure [16, 34-36, 61], which provide the rare examples exhibiting field-induced metamagnetic quantum criticalities similar to the observation in $Nd_3BWO_9$. But, these two are quite distinct, as the former can be tuned by the RKKY interaction and Kondo coupling, whereas the above two kinds of interactions are completely absent in $Nd_3BWO_9$ due to its insulating nature. Based on this consideration, the anisotropic multi-critical behaviors should be related to the strong quantum spin fluctuation driven by frustration with an anisotropic form, and indeed the low-*T* fitted $\theta_{CW}$ = -0.97 K (*B* // c) and $\theta_{CW}$= -1.31 K (*B* // a*) support the exchange anisotropy with magnetic frustration index $f=\left|\theta_{CW}/T_N\right|$ = 3.0 (*B* // *c*) and $f = 4.1$ (*B* // *a**), respectively. Moreover, the nearest-neighboring (NN) dipolar interaction (D) and magnetic exchange coupling ($J_{nn}$) can be evaluated by the following relations [5,20]: $D = \frac{-\mu_0\mu_{eff}^2}{4\pi r_{nn}^3}$ and $\theta_{CW}=-\frac{zJ_{nn}S_{eff}(S_{eff}+1)}{3k_B}$, where $\mu_{eff}$ is the low-temperature fitted effective moment, $r_{nn}$ is the NN distance of Nd ions and z is the number of the nearest neighbour spins, respectively. The obtained interplane dipolar interaction $D_{inter}$= 0.08 K and exchange interaction $J_{nn}^{inter}=-0.65\text{ K}$ alongside the intraplane dipolar interaction $D_{intra}$=0.06 K and the exchange interaction $J_{nn}^{intra}=-1.31\text{ K}$ support the existence of anisotropic magnetic interactions along different axes. From the structure viewpoint, the AB-type kagome-lattice topology of $Nd^{3+}$ moments and their magnetic exchange pathways can account for this anisotropy, where the nearest neighboring Nd-Nd ions form the zigzag chain structure along the *c*-axis and the breathing kagome-lattice connections within the *ab*-plane. For the above two types of spin-lattice topology, since each one can independently generate quantum spin fluctuation, their intertwining will consequently generate the 3D anisotropic exchange interaction as well as quantum spin fluctuation. Further considering the low site symmetry around $Nd^{3+}$ ions [37, 62], the Nd-O-Nd exchange pathways exhibit different bond distances and bond angles within or perpendicular to the *ab*-plane, also contribute to the magnetic anisotropy.

For the metamagnetic bicriticality, the phase diagram analysis on RE-based frustrated heavy systems have revealed that both kondo effect and spin frustration can be effective parameters on tuning the location of metamagnetic CPs [34, 63], which depends on their strengths. According to that scenario, for *B* // *c*, the tuning parameter of QBCP in $Nd_3BWO_9$ can be related the quantum fluctuation by a moderate frustration strength with $f = 3.0$, then by reducing and enhancing the strength of magnetic frustration, it is expected to generate a finite-temperature BCP as observed in anisotropic AFMs without apparent magnetic frustration and metamagnetic spin disorder state existing at finite interval field regimes. For the former situation, a finite-temperature critical end point associated with a first-order quantum transition lines was recently observed in the frustrated shastry-sutherland-lattice and pyrochlore-lattice magnets [64,65], that terminates a first-order metamagnetic transition line. For the latter case, field induced intermediate quantum spin liquid phase at finite interval fields has been proposed in kagome systems [16,34,36,66], which call for further experimental study in the future.

Compared to *B* // *c*, a tetracritical phase diagram is realized for *B* // *a\**. For the appearance of tetracritical point, it was initially predicted in the anisotropic 3D Heisenberg AFMs as well as randomly mixed magnets [52,53, 67]. In those systems, distinct to the 1st critical line separating two ordered phases below the BCP in the bicritical phase diagram, an intermediate magnetic phase can be stabilized which exhibit the simultaneous ordering of two spin components (diagonal and off-diagonal moments) below the TP, and which phase can be called a spin analog of supersolid state due to the simultaneous breaking of discrete lattice translation symmetry and spin rotational symmetry [48,49]. As one typical spin-system, a two-dimensional uniaxial AFMs with cubic anisotropy [68], the appearance of tetracritical point refers to a situation that two types of critical lines encompassing distinct order parameters (Ising line and XY Kosterlitz-Thouless line) intersect at finite temperature, that will enclose a novel intermediate phase. Here, the easy *c*-axis magnetization and the accompanied in-plane six-fold magnetic anisotropy support the $Nd_3BWO_9$ to be a 3D AFM with Ising anisotropy. Along the *a\**-axis, the observation of $1/6M_S$ in *M*(B) curves corresponds to this novel intermediate phase possibly stabilized by the intraplane breathing kagome-lattice, even its nature and related spin configurations still need to be clarified. In light of this, the single crystal neutron scattering at low temperatures is required. More interestingly, to explore the potential novel magnetic states and multicritical points, another important investigation is the synthetic tunability on the spin frustration strength through the combination of magnetic field and other nonthermal parameters including chemical doping, pressure as well as uniaxial strain.

In summary, the distorted kagome antiferromagnet $Nd_3BWO_9$ give an experimental access to investigate the field-induced quantum phase transition. Below $T_N$~0.32 K, our isothermal magnetizations reveal the different metamagnetic crossover behaviors with the observation of a 1/3 fractional magnetization for $B // c$ and two successive fractional magnetization ($1/6M_S$, $1/4M_S$) anomalies for $B // a^*$, respectively. In combination with the thermodynamic results, the constructed $B$-$T$ magnetic phase diagrams establish the anisotropic multicritical behaviors with the detection of metamagnetic bicritical point ($B // c$) and tetracritical point ($B // a^*$). The above field-induced quantum criticalities in $Nd_3BWO_9$ open up the possibility to explore the multiple QCs tuned by the nonthermal parameters.

## ACKNOWLEDGMENTS

We thank Wei Li and Gang Chen for the valuable discussions. We acknowledge financial support from the National Natural Science Foundation of China (Grant Nos. 11874158, and U20A2073), the Fundamental Research Funds of Guangdong Province (Grant No. 2022A1515010658) and Guangdong Basic and Applied Basic Research Foundation (Grant No.2022B1515120020). This work was supported by the synergetic extreme condition user facility (SECUF), and a portion of magnetic measurement was performed on the Steady High Magnetic Field Facilities. We would like to thank the staff of the analysis center of Huazhong University of Science and Technology for their assistance in structural characterizations.

**Figure captions**

FIG. 1 Crystal structure and elementary magnetic properties of $Nd_3BWO_9$ single crystal. (a) The schematic view of alternating-stacked Nd-kagome layers along the *c*-axis. (b) The distorted kagome net of $Nd^{3+}$ ions within the *ab*-plane. (c) The local oxygen environment of $Nd^{3+}$ ions. (d) A typical as-grown single crystal of $Nd_3BWO_9$. (e) The isothermal magnetization *M*(*B*) curves at *T* = 2 K for *B* along the *c*, *a*\* and *b*\* axis, the inset show the definition of the $c$, $a^*$ and $b^*$ axes. (f) The angle-dependent magnetization for $B$ rotated within the $ab$-plane at $T$ = 2 K and $B$ = 7 T, the $a^*$-axis and $b^*$-axis is along the direction with maximal and minimal magnetization values, respectively. (g) The inverse magnetic susceptibility $1/\chi(T)$ measured at $B$ = 1T over temperature range of 0.4 - 300 K for $B // c$ and $B // a^*$. Inset shows the low temperature part of $1/\chi(T)$, the solid lines show the Curie-Weiss fits in the temperature range 2-10 K.

FIG.2 (a) Zero-field specific heat $C_p(T)$ of Nd$_3$BWO$_9$ (Solid cycles) and La$_3$BWO$_9$ (green line), the dashed gray line shows the estimated contribution from nuclear Schottky effect. The $T_N$ and $T_{sr}$ is marked by the arrow and vertical line, respectively. (b) Zero-field magnetic specific heat $C_M(T)$ and associated magnetic entropy $S_M(T)$ of Nd$_3$BWO$_9$. (c) Low temperature specific heat $C_p(T)$ at different magnetic fields along $c$-axis, each data set is offset by 2 J/mol K$^2$ for clarification, the dashed blue and gray lines guide the evolution of $T_N$ and $T_{sr}$ in $C_p(T)$ under magnetic field.

FIG.3 The isothermal field-dependent magnetization $M(B)$ curves of Nd$_3$BWO$_9$ single crystal at low temperatures (0.05 K $\leq T \leq$ 2 K) for (a) $B // c$, (b) $B // a^*$ and (c) $B // b^*$, respectively. The normalized magnetization ($M/M_S$) and the derivative of magnetization ($dM/dB$) curves at 50 mK for (d) $B // c$, (e) $B // a^*$ and (f) $B // b^*$, respectively.

FIG.4 The low-temperature magnetization $M(T)$ curves of Nd$_3$BWO$_9$ single crystal at different fields for (a) $B // c$ and (b) $B // a^*$, the arrow and solid cycles indicate the positions of $T_N$ and $T_{sr}$ in $M(T)$ curves.

FIG. 5 Field dependence of specific heat $C_M(B)$ of Nd$_3$BWO$_9$ single crystal at selected temperatures (a) $T$< 0.2 K and (b) $T$ >0.3 K for $B // c$, (c) $T$ < 0.2 K and (d) $T$ > 0.3 K under $B // a^*$. The solid cycles and open stars indicate the field positions for the peaks of $C_M(B)$ curves, where $B_{c1}$, $B_{c2}$ and $B_{a^*1}$, $B_{a^*2}$ are the metamagnetic critical fields and $B_{s1,c}$, $B_{s2,c}$ and $B_{s1,a^*}$, $B_{s2,a^*}$ indicate the critical fields of magnetic crossover at high fields.

FIG. 6 The magnetic phase diagrams of Nd$_3$BWO$_9$ single crystal. (a) The $B$-$T$ phase diagram for $B // c$, the black and dashed gray lines show the fits of phase boundaries by power law, the quantum bicritical point (QBCP) and tricritical point (TCP) are denoted by the black ball and pink ball, respectively. (b) The $B$-$T$ phase diagram for $B // a^*$, the tetracritical point (TP) and quantum critical point (QCP) are denoted by the orange ball and black ball, respectively. The points on the phase boundaries are determined by magnetic susceptibility, magnetization and specific heat results.

FIG. 7 (a) Magnetocaloric effect $T(B)$ results at different temperatures measured in the adiabatic condition and (b) the contour plot of $B$-$T$ phase diagrams using the calculated $\Gamma_{mag}(B)$ from the $T(B)$ data for $B // c$. (c) The $T(B)$ data and (d) contour plot of $B$-$T$ phase diagrams using $\Gamma_{mag}(B)$ results for $B // a^*$. In (a) and (c), the arrows guide the evolution of phase boundaries determined by the peaks of $dT/dB$ near the metamagnetic crossover, the dashed lines indicate the field positions of CPs. In (b) and (d), the phase boundaries are determined by magnetocaloric effect and specific heat, the dashed lines represent the power-law fits to the phase boundary lines.

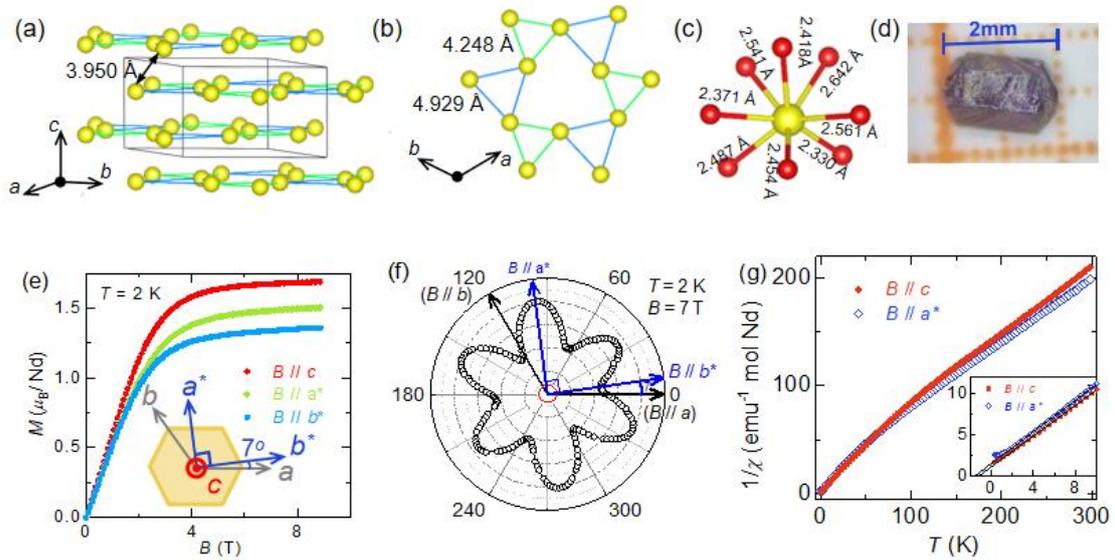

FIG. 1

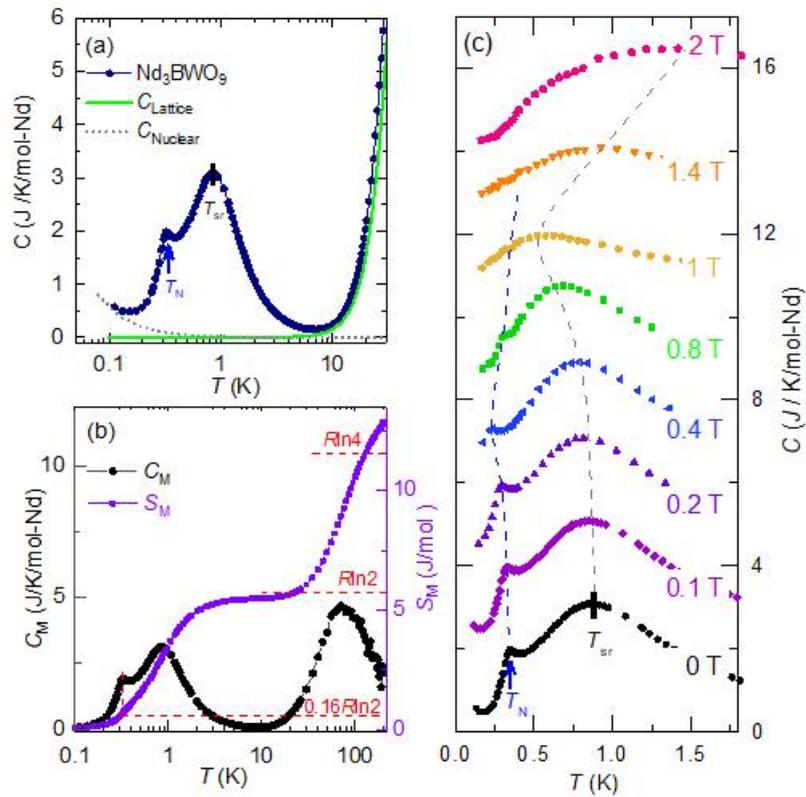

FIG. 2

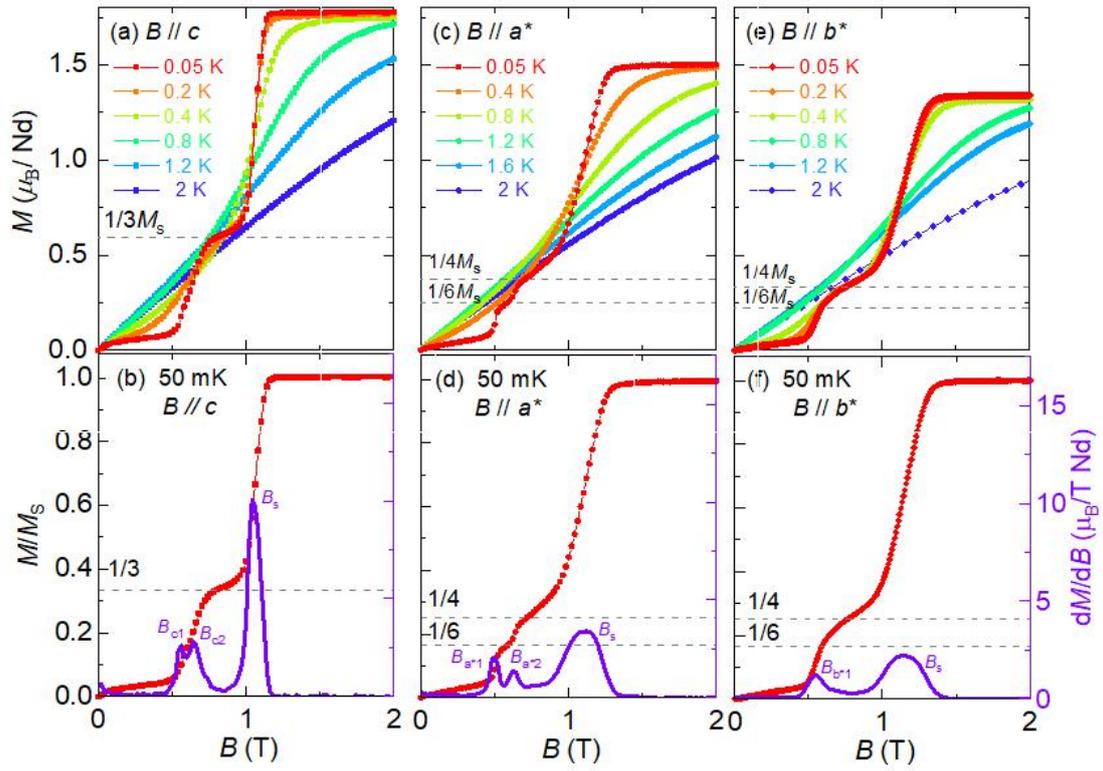

FIG.3

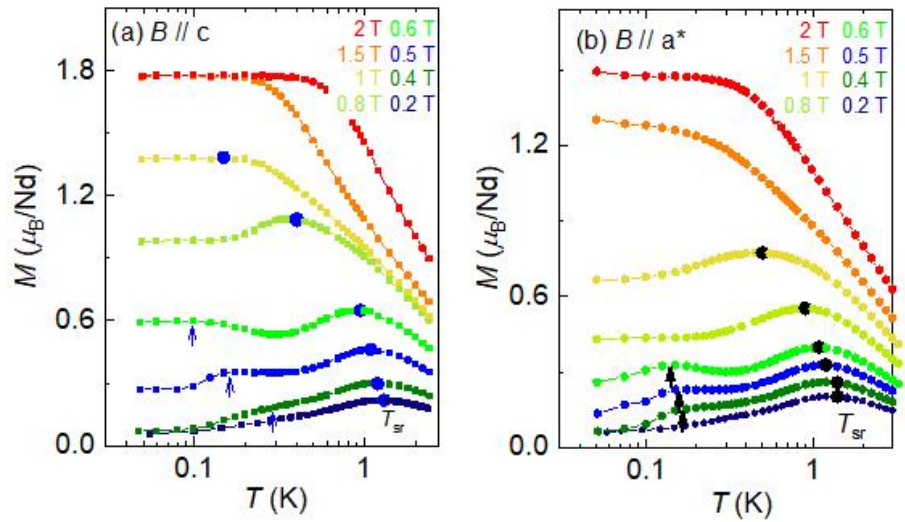

FIG.4

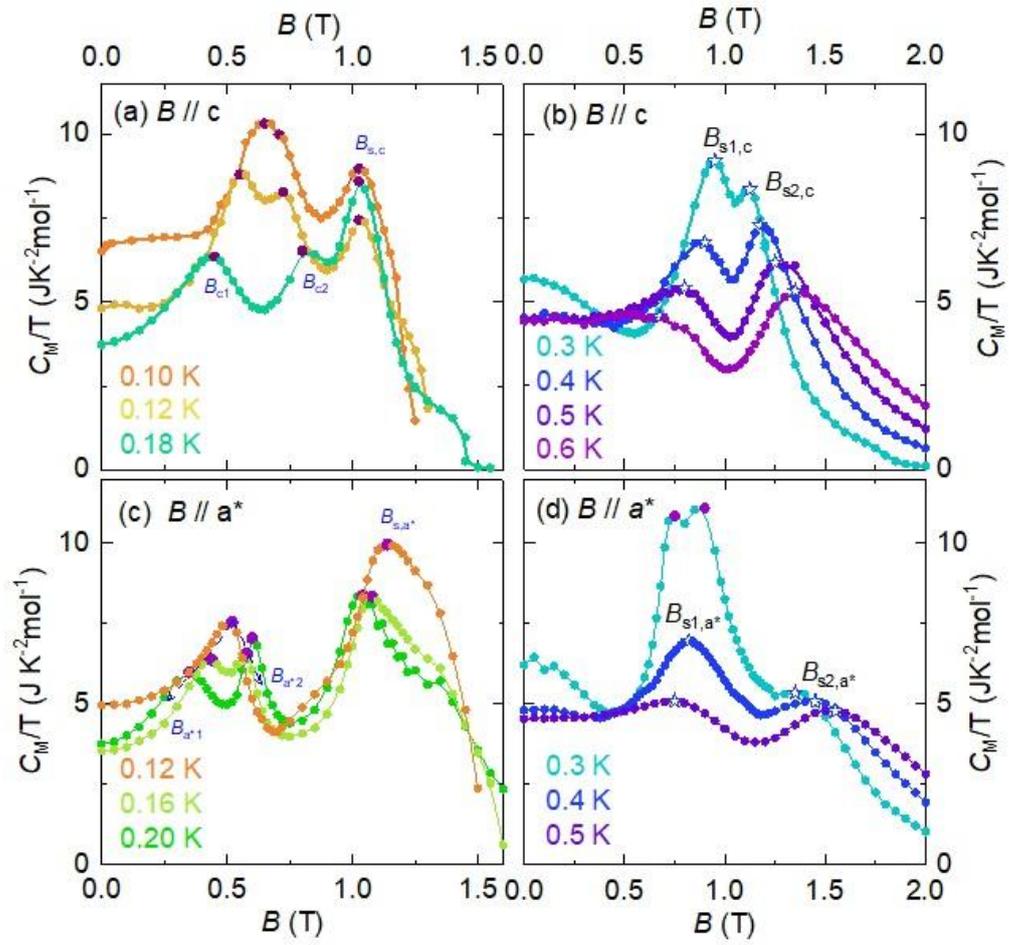

FIG. 5

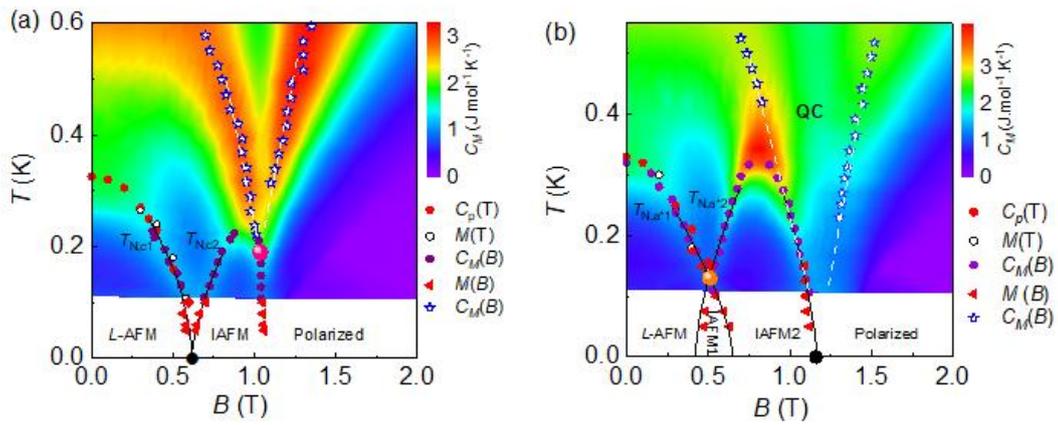

FIG. 6

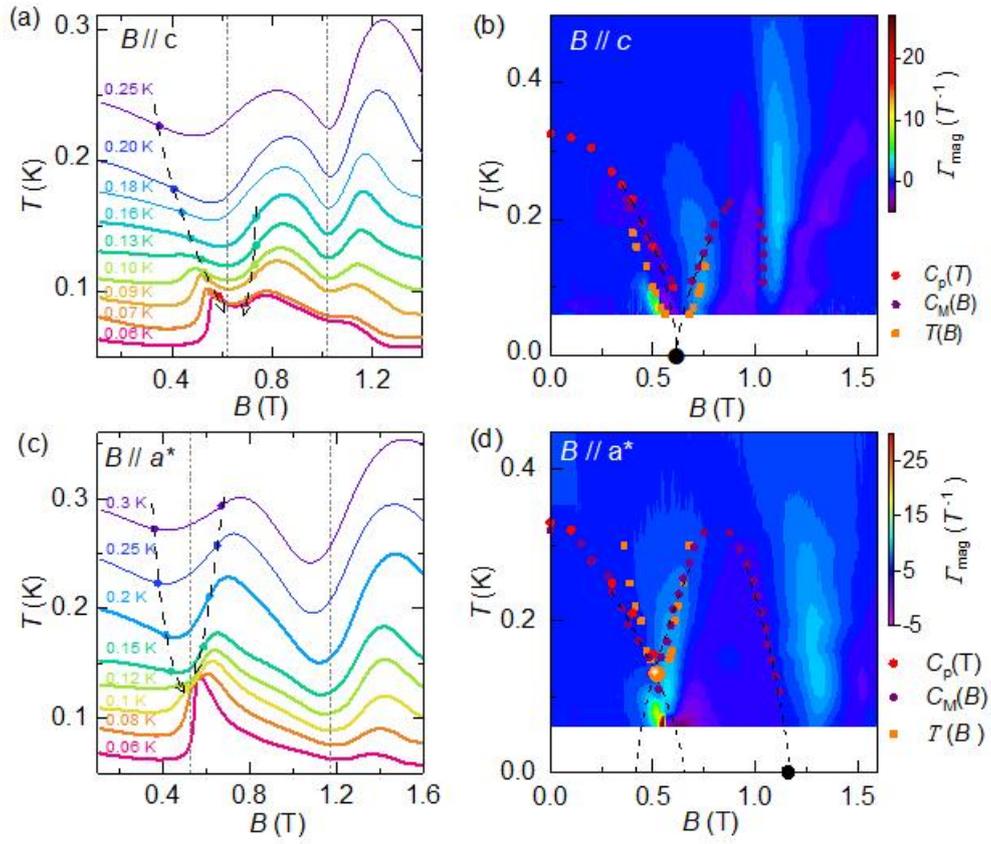

FIG. 7